\documentclass[conference,a4paper]{IEEEtran}

\usepackage[utf8]{inputenc} 
\usepackage[T1]{fontenc}
\usepackage{url}
\usepackage{ifthen}
\usepackage{cite}
\usepackage[cmex10]{amsmath} 
\usepackage{amssymb}

\newtheorem{thm}{Theorem}[section]

\newtheorem{cor}[thm]{Corollary}

\newtheorem{exmp}[thm]{Example}

\begin{document}
\title{Optimal Error Correcting Delivery Scheme for an Optimal Coded Caching Scheme with Small Buffers} 

\author{%
  \IEEEauthorblockN{Nujoom Sageer Karat, Anoop Thomas and B. Sundar Rajan}
  \IEEEauthorblockA{Department of Electrical Communication Engineering, Indian Institute of Science, Bengaluru 560012, KA, India \\
E-mail: \{nujoom,thomas,bsrajan\}@iisc.ac.in}
}

\maketitle

\begin{abstract}
	
Optimal delivery scheme for coded caching problems with small buffer sizes and the number of users no less than the amount of files in the server was proposed by Chen, Fan and Letaief [``Fundamental limits of caching: improved bounds for users with small buffers," \textit{IET Communications}, 2016]. This scheme is referred to as the CFL scheme. In this paper, the link between the server and the users is assumed to be error prone only during the delivery phase. Closed form expressions for average rate and peak rate of error correcting delivery scheme for CFL prefetching scheme is obtained. An optimal error correcting delivery scheme for caching problems employing CFL prefetching is proposed.

\end{abstract}
\section{INTRODUCTION}
The problem of coded caching introduced in \cite{MaN}, plays a crucial role in reducing peak hour traffic in networks. A part of the content is made available in local cache of users so that traffic can be reduced at peak hours. Coded caching scheme involves two phases: a placement phase and a delivery phase. In the placement phase or the prefetching phase, which is performed during off-peak times, the entire database is made available to each user. Users fill their cache with the available data. Delivery phase is performed during peak traffic time. During placement phase some parts of files have to be judiciously cached at each user in such a way that the rate of transmission is reduced during the delivery phase. The prefetching can be done with or without coding. If during prefetching, no coding of parts of files is done, the prefetching scheme is referred to as uncoded prefetching \cite{MaN, YMA}. If coding is done during prefetching stage, then the prefetching scheme is referred to as coded prefetching \cite{CFL,JV}.

The seminal work in \cite{MaN} shows that apart from the \textit{local caching gains} obtained by placing contents at user caches before the demands are revealed, a \textit{global caching gain} can be obtained by coded transmissions. This scheme is extended to decentralized scheme in \cite{MaN2}. More extensions to non-uniform demands \cite{NiM} and online coded caching \cite{PMN} are also available in literature. 

If the shared bottleneck link between the server and the users is error-prone during the delivery phase, an error correcting delivery scheme is required. The minimum average rate and minimum peak rate of error correcting delivery schemes is characterized in \cite{KTR}. The placement phase is assumed to be error-free. This assumption can be justified as during placement phase there is no bandwidth constraint and any number of re-transmissions can be done to make the placement error-free. A similar model in which the delivery phase takes place over a packet erasure broadcast channel was considered in \cite{BWT}.

In this paper, we consider the coded caching problem considered in \cite{CFL}, where coded  prefetching involves coding of parts of files. Optimal error correcting delivery scheme is proposed for this scheme. The main contributions of this paper are as follows.

\begin{itemize}
	\item The coded caching scheme proposed in \cite{CFL} is considered for the case where the number of users is the same as the number of files. The minimum number of transmissions required for correcting finite number of transmission errors is obtained for this case (Section \ref{Sec:NequaltoK} ).
	\item When the number of users is greater than the number of files, a different prefetching scheme is employed in \cite{CFL}. For this caching strategy, the minimum number of transmissions required for correcting finite number of transmission  errors is obtained (Section \ref{Sec:KgreaterthanN}).
 	\item An error correcting delivery scheme for coded caching problem with coded prefetching for small buffer sizes is proposed. We find expressions for average rate and peak rate of this error correcting delivery scheme (Section \ref{Sec:ECDforCFL}). 
\end{itemize}

In this paper $\mathbb{F}_{q}$ denotes the finite field with $q$ elements, where $q$ is a power of a prime, and $\mathbb{F}^{*}_{q}$ denotes the  set of all non-zero elements of $\mathbb{F}_q$. The notation $[K]$ is used for the set $\{1,2, \ldots,K\}$ for any integer $K$. For a $K \times N$ matrix $L$, $L_i$ denotes its $i$th row. For vector spaces $U,V$, $U < V $ denotes that $U$ is a subspace of $V$.

A linear $[n,k,d]_q$ code $\mathcal{C}$ over $\mathbb{F}_q$ is a $k$-dimensional subspace of $\mathbb{F}^{n}_{q}$ with minimum Hamming distance $d$. The vectors in $\mathcal{C}$ are called codewords. A matrix ${G}$ of size $k \times n$ whose rows are linearly independent codewords of $\mathcal{C}$ is called a generator matrix of $\mathcal{C}$. A linear $[n,k,d]_q$ code $\mathcal{C}$ can thus be represented using its generator matrix ${G}$ as,
$ \mathcal{C} = \{ {\mathbf{y}}G: {\mathbf{y}} \in \mathbb{F}^{k}_{q} \}.$ 
Let $N_{q}[k,d]$ denote the length of the shortest linear code over $\mathbb{F}_q$ which has dimension $k$ and minimum distance $d$.

%%%%%%%%%%%%%%%%%%%%%%%%%%%%%%%%%%%%%%%%%%%%%%%%%%%%%%%%%%%%%%%%%%%%%
\section{Preliminaries and Background}
To obtain error correcting delivery schemes we use results from error correction for index coding with coded side information. In this section we recall results from error correction for index coding with coded side information introduced in \cite{ByC}. We also review the coded caching scheme with coded prefetching proposed in \cite{CFL}. 

%%%%%%%%%%%%%%%%%%%%%%%
\subsection{Generalized Index Coding Problem and Error Correction}
The index coding (IC) problem with side-information was introduced by Birk and Kol \cite{BiKol, BiK}. A sender broadcasts messages through a noiseless shared channel to multiple receivers, each demanding certain messages and knowing some other messages a priori as side-information. The sender needs to meet the demands of each receiver in minimum number of transmissions. In \cite{MiD} and \cite{SDC}, a generalization of the index coding problem was discussed, where the demands of the receivers and the side-information are linear combinations of the messages. In \cite{SDC}, the authors refer to this class of problems as Generalized Index Coding (GIC) problems.

An instance $\mathcal{I}$ of GIC problem is described formally as follows. There is a message vector $X={x}^T = (x_1,x_2, \ldots, x_n)^T \in \mathbb{F}_q^{n \times 1}$ and there are $m$ receivers. The $i$th receiver demands a linear combination of the messages $R_iX$, for some $R_i \in \mathbb{F}_q^{1 \times n}$, where $R_i$ is the request vector and $R_iX$ is the request packet of the $i$th receiver. The side-information is represented by a matrix $V^{(i)} \in \mathbb{F}_q^{s_i \times n}$, where $s_i$ is the number of packets possessed as side-information by the $i$th receiver. Though $X$ is unknown to the receiver $i$, it can generate any vector in the row space of $V^{(i)}$, denoted by $\mathcal{X}^{(i)}$. Let $R$ be an $m \times n$ matrix over $\mathbb{F}_q$ having $R_i$ as its $i$th row. The matrix $R$ represents the demands of all the $m$ receivers. In the definition of GIC problem in \cite{SDC} the source is assumed to possess only certain linear combinations of messages. In our work, it is assumed that all the messages are independent and the source possesses all of them.

The min-rank of an instance $\mathcal{I}$ of the GIC problem over $\mathbb{F}_q$ is defined as
$$ \kappa(\mathcal{I}) = \text{min}\{\text{rank}(A+R): A\in \mathbb{F}_q^{m \times n}, A_i \in \mathcal{X}^{(i)}, i \in [m] \}.$$ It is shown in \cite{ByC} that the min-rank is optimal length of linear generalized index code. For each $i \in [m]$, the set $\mathcal{Z}^{(i)}$ is defined as 
\begin{equation} 
 \mathcal{Z}^{(i)} \triangleq \{Z \in \mathbb{F}_q^{n \times 1} : V^{(i)}Z=0, R_iZ \neq 0 \}. \label{eq:setz}
\end{equation} 
The set $\mathcal{J}(\mathcal{I})$ is defined as $  \mathcal{J}(\mathcal{I}) \triangleq \{U<\mathbb{F}_q^n: U \setminus\{0 \} \subset \cup_{i \in [m]}\mathcal{Z}^{(i)} \} .$  The maximum dimension of any element of $\mathcal{J}(\mathcal{I})$ is called the generalized independence number, denoted by $\alpha(\mathcal{I})$. Thus dimension of any subspace of $\mathbb{F}_q^n$ in $\cup_{i \in [m]}\mathcal{Z}^{(i)} \cup \{0\}$ serves as a lower bound for $\alpha(\mathcal{I})$. It was shown in \cite{ByC} that the min-rank serves as an upper bound for the generalized independence number, 
\begin{equation}
\alpha(\mathcal{I}) \leq \kappa(\mathcal{I}).
\label{eq:alleqka}
\end{equation}

Generalized index coding problems were classified in \cite{KSR}. In Generalized Index Coding with Coded Side-Information GIC (CSI) problems,  demand of every receiver is uncoded but the side-information is coded. In Generalized Index Coding with Coded Demands GIC (CD) problems, the side-information of every receiver is uncoded but the demand is coded. In our work the focus is on GIC (CSI) problems.

Error correcting index codes were introduced in \cite{DSC} and later extended for generalized index coding problems in \cite{ByC}. An Error Correcting Generalized Index Code (ECGIC) is a map that encodes the message vector $X$ such that each user, given its side-information and received transmissions with at most $\delta$ transmission errors, can decode its requested packet $R_iX \in \mathbb{F}_q$. An optimal linear $(\mathcal{I}, \delta)$-ECGIC over $\mathbb{F}_q$ is a linear $(\mathcal{I}, \delta)$-ECGIC over $\mathbb{F}_q$ of the smallest possible length $\mathcal{N}_{q}[\mathcal{I},\delta]$. The length of an optimal linear $(\mathcal{I}, \delta)$-ECGIC, $\mathcal{N}_{q}[\mathcal{I},\delta]$ satisfies
\begin{equation}
\underbrace{N_q[\alpha(\mathcal{I}), 2\delta + 1]~ \leq ~}_{\alpha\text{-bound}}  \mathcal{N}_{q}[\mathcal{I},\delta] \underbrace{~\leq~ N_q[\kappa(\mathcal{I}), 2\delta + 1]}_{\kappa\text{-bound}}. \label{eq:bds}
\end{equation}
where $N_{q}[k,d]$ is the length of an optimal linear classical error-correcting code of dimension $k$ and minimum distance $d$ over $\mathbb{F}_{q}$ \cite{DSC, ByC}. 

The $\kappa$-bound is obtained by concatenating an optimal linear classical error correcting code and an optimal linear index code. Thus for any index coding problem, if $\alpha (\mathcal{I})$ is same as $\kappa_q(\mathcal{I})$, then concatenation scheme would give optimal error correcting index codes \cite{SaR,SagR,KSR}. 
%%%%%%%%%%%%%%%%%%%%%%%
\subsection{Error Correcting Coded Caching Scheme}
Error Correcting coded caching scheme was proposed in \cite{KTR}. The server is connected to $K$ users through a shared link which is error prone. 
The server has access to $N$ files $X_1, X_2, \ldots, X_N$, each of size $F$ bits. Every user has an isolated cache with memory $MF$ bits, where $M \in [0,N]$.  A prefetching scheme is denoted by ${\mathcal{M}}$. During the delivery phase, only the server has access to the database. Every user demands one of the $N$ files. The demand vector is denoted by $\mathbf{d} = (d_1, \ldots, d_K)$, where $d_i$ is the index of the file demanded by user $i$. The number of distinct files requested in $\mathbf{d}$ is denoted by $N_e(\mathbf{d})$. The set of all possible demands is denoted by $\mathcal{D} = \{1, \ldots, N\}^K.$ During the delivery phase, the server informed of the demand $\mathbf{d}$, transmits a function of $X_1, \ldots, X_N$, over a shared link. Using the cache contents and the transmitted data, each user $i$ needs to reconstruct the requested file $X_{d_{i}}$ even if $\delta$ transmissions are in error.

For the $\delta$-error correcting coded caching problem, a communication rate $R(\delta)$ is \textit{achievable} for demand $\mathbf{d}$ if and only if there exists a transmission of $R(\delta)F$ bits such that every user $i$ is able to recover its desired file $X_{d_{i}}$ even after at most $\delta$ transmissions are in error. Rate  $R^*(\mathbf{d}, \mathcal{M}, \delta)$ is the minimum achievable rate for a given $\mathbf{d}$, $\mathcal{M}$ and $\delta$. The average rate $R^*(\mathcal{M}, \delta)$ is defined as the expected minimum average rate given $\mathcal{M}$ and $\delta$ under uniformly random demand. Thus $ R^*(\mathcal{M}, \delta) = \mathbb{E}_{\mathbf{d}}[R^*(\mathbf{d}, \mathcal{M}, \delta)].$

The average rate depends on the prefetching scheme $\mathcal{M}$. The minimum average rate  $ R^*(\delta)= \min_{\mathcal{M}} R^*(\mathcal{M}, \delta)$  is the minimum rate of the delivery scheme over all possible $\mathcal{M}$. The rate-memory trade-off for average rate is finding the minimum average rate $R^*(\delta)$ for different memory constraints $M$. Another quantity of interest is the peak rate, denoted by $R^*_{\text{worst}}(\mathcal{M}, \delta)$, which is defined as
$R^*_{\text{worst}}(\mathcal{M}, \delta) = \max_{\mathbf{d}} R^*(\mathbf{d}, \mathcal{M}, \delta).$ 
The minimum peak rate is defined as
$ R^*_{\text{worst}}(\delta)= \min_{\mathcal{M}} R^*_{\text{worst}}(\mathcal{M}, \delta).$

%%%%%%%%%%%%%%%%%%%%%%%
\subsection{Coded Caching Scheme with Coded Prefetching}
A coded caching scheme for small cache sizes involving coded prefetching was proposed in \cite{CFL}. We call this scheme as Chen Fan Letaief (CFL) scheme. The system consists of a server and $K$ users. The server has access to $N$ files  $X_1, X_2, ..., X_N$, each of size $F$ bits. Every user has an isolated cache with memory $MF$ bits.  The prefetching scheme is denoted by $\mathcal{M}_{\text{CFL}}$. 

\begin{itemize}
	\item Consider the case when $N=K$ and $M=1/K$. Each file is split into $N$ subfiles, i.e., $X_i=(X_{i,1}, X_{i,2}, \ldots, X_{i,N}).$ During prefetching, the cache of user $j$ is designed to be $Y_j= X_{1,j} \oplus X_{2,j} \oplus \ldots \oplus X_{N,j}$, an XORed version of subfiles. It is shown in \cite{CFL} that $R(\delta = 0)=N_e(\mathbf{d})$ for $N_e(\mathbf{d}) \leq N-1$ and $R(\delta=0)=N-1$ for $N_e(\mathbf{d})=N$ are achievable. Furthermore if $M \in [0,1/N],$ $R(M,\delta = 0) = N(1-M)$ is achievable by memory sharing.
	\item Consider $K>N$ and $M=1/K.$ Each file is split into $NK$ subfiles, i.e., $X_i=(X_{i,1}, X_{i,2}, \ldots, X_{i,NK}).$ The cache of user $i$ is given by $Y_i=X_{1,N(i-1)+j} \oplus \ldots \oplus X_{N,N(i-1)+j},$ for $j=1, 2, \ldots, N.$ For the number of distinct demands $N_e(\mathbf{d}) \leq N-1$ files, it is shown in \cite{CFL} that $R (\delta=0)= N_e(\mathbf{d})$ is achievable. For $N_e(\mathbf{d})=N$, the rate $R(\delta=0)= N-N/K$ is achievable. Furthermore if  $M \in [0,1/K],$ $R(M, \delta=0) = N(1-M)$ is achievable by memory sharing.
\end{itemize} 

For a fixed prefetching $\mathcal{M}$ and for a fixed demand $\mathbf{d}$, the delivery phase of a coded caching problem is an index coding problem \cite{MaN}. In fact, for fixed prefetching, a coded caching scheme consists of $N^K$  parallel index coding problems one for each of the $N^K$ possible user demands. Thus finding the minimum achievable rate for a given demand $\mathbf{d}$ is equivalent to finding the min-rank of the equivalent index coding problem induced by the demand $\mathbf{d}$.

Consider the CFL prefetching scheme $\mathcal{M}_{\text{CFL}}$. The index coding problem induced by the demand $\mathbf{d}$  for CFL prefetching is denoted by $\mathcal{I}(\mathcal{M}_{\text{CFL}}, \mathbf{d}).$ Each subfile $X_{i,j}$ corresponds to a message in the index coding problem. Since prefetching is coded, $\mathcal{I}(\mathcal{M}_{\text{CFL}}, \mathbf{d})$ represents a GIC (CSI) problem.

%%%%%%%%%%%%%%%%%%%%%%%%%%%%%%%%%%%%%%%%%%%%%%%%%%%%%%%%%%%%%%%%%%%%%
\section{Generalized Independence Number for $\mathcal{I}(\mathcal{M}_{\text{CFL}}, \mathbf{d})$ }

In this section we find a closed form expression for generalized independence number $\alpha(\mathcal{M}_{\text{CFL}}, \mathbf{d})$ of the index coding problem $\mathcal{I}(\mathcal{M}_{\text{CFL}}, \mathbf{d})$. There are two different prefetching schemes employed in \cite{CFL} depending upon the relationship between number of messages and number of receivers. For both these prefetching schemes, the generalized independence number of the corresponding index coding problem is shown to be equal to the min-rank.

\subsection{Number of files equal to number of users ($N=K$)}
\label{Sec:NequaltoK}

In the CFL prefetching scheme, each file is split into $N$ subfiles. Hence the number of messages in $\mathcal{I}(\mathcal{M}_{\text{CFL}}, \mathbf{d})$ is $N^2$. Each user is split into $N$ receivers each demanding one message. Hence there are a total of $N^2$ receivers. From the expressions of the achievable rates in \cite{CFL}, we get the min-rank $\kappa(\mathcal{M}_{\text{CFL}}, \mathbf{d})$ as
\begin{equation}
\kappa(\mathcal{M}_{\text{CFL}}, \mathbf{d}) \leq 
\begin{cases}
N (N_e(\mathbf{d}))              & \text{if} \  N_e(\mathbf{d}) \leq N-1\\
N(N-1)             & \text{if} ~ N_e(\mathbf{d}) = N
\end{cases}.
\label{eq: kappanek}
\end{equation}
We find the generalized independence number $\alpha(\mathcal{M}_{\text{CFL}}, \mathbf{d})$ for $\mathcal{I}(\mathcal{M}_{\text{CFL}}, \mathbf{d})$. The technique of obtaining $\alpha(\mathcal{M}_{\text{CFL}}, \mathbf{d})$  is illustrated in the following example.

\begin{exmp}
Consider a coded caching problem with    $N=K=3$, $M=1/3$. Since $M=1/K$, the CFL scheme is used for solving the coded caching problem. Each file is split into $N=3$ subfiles as 
	$ X_1=(X_{1,1}, X_{1,2}, X_{1,3})$,  
	$ X_2=(X_{2,1}, X_{2,2}, X_{2,3})$ and 	$ X_3=(X_{3,1}, X_{3,2}, X_{3,3})$. Let  $X=(X_{1,1},X_{1,2}, X_{1,3},X_{2,1},X_{2,2},X_{2,3},X_{3,1},X_{3,2},X_{3,3})$ denote the vector obtained by concatenating $X_1, X_2$ and $X_3$.
The cache contents of user $i$ is $Y_i=(X_{1,i} \oplus X_{2,i} \oplus X_{3,i})$ for $i=1,2,3.$ 

For a given demand $\mathbf{d}$, this problem becomes the generalized index coding problem $\mathcal{I}(\mathcal{M}_{\text{CFL}}, \mathbf{d})$. We calculate the generalized independence number $\alpha(\mathcal{M}_{\text{CFL}}, \mathbf{d})$ for this problem. For different demands,  generalized independence number is calculated and it is shown to be equal to min-rank of the corresponding generalized index coding problem. 

First consider that all the demands are distinct, i.e., $N_e(\mathbf{d})=3$. Without loss of generality we can assume that the demand is $\mathbf{d}=(1,2,3).$
Consider the equations
	$e_1: X_{1,1} \oplus X_{2,1} \oplus X_{3,1}=0,$
	$e_2: X_{1,2} \oplus X_{2,2} \oplus X_{3,2}=0 \text{ and }$
	$e_3: X_{1,3} \oplus X_{2,3} \oplus X_{3,3}=0.$
	Let $\mathcal{S}$ be the subspace of $\mathbb{F}_q^9$, which consists of the vectors satisfying the equations $e_1,e_2$ and $e_3$. From rank nullity theorem, dim$(\mathcal{S}) \geq 6$. The induced generalized index coding problem $\mathcal{I}(\mathcal{M}_{\text{CFL}}, \mathbf{d})$ has 9 messages and 9 receivers. For this case, \eqref{eq:setz} can be rewritten as
$ \mathcal{Z}^{(i,j)} \triangleq \{Z \in \mathbb{F}_q^{9} : e_i, X_{d_i,j} \neq 0 \}.$ Let $A= \cup_{i,j \in [3]}\mathcal{Z}^{(i,j)} \cup \{\mathbf{0}\}$. The generalized independence number is the maximum dimension of any subspace of $\mathbb{F}_q^{9}$ in $A$. We claim that all the vectors  of $\mathcal{S}$ belong to the set $A$. This would mean $\alpha(\mathcal{M}_{\text{CFL}}, \mathbf{d}) \geq \text{dim}(\mathcal{S}) \geq 6$. From the definition of $A$, it is clear that the all zero vector $\mathbf{0}$ belonging to $\mathcal{S}$ also belongs to $A$. Any other vector in $\mathcal{S}$ will have at least one non-zero coordinate $X_{i,j}$. The vector belonging to $\mathcal{S}$, having $X_{i,j} \neq 0$ belongs to $\mathcal{Z}^{(i,j)}$. Thus all vectors in $\mathcal{S}$ lie in $A$ and $\alpha(\mathcal{M}_{\text{CFL}}, \mathbf{d}) \geq 6.$ From \eqref{eq: kappanek}, we get $\kappa(\mathcal{M}_{\text{CFL}}, \mathbf{d}) \leq 6$. Hence by \eqref{eq:alleqka}, we have $\alpha(\mathcal{M}_{\text{CFL}}, \mathbf{d})=\kappa(\mathcal{M}_{\text{CFL}},\mathbf{d})=6.$

Consider the case when $N_e(\mathbf{d})=2$.  Let $\mathbf{d}=(1,2,1).$ Here we consider the same set of equations $e_1,e_2$ and $e_3$ and their solution space $\mathcal{S}$. Following the same argument as before, any vector in $\mathcal{S}$ with $X_{i,j}\neq 0$ for $i=1,2$ and $j=1,2,3$ lies in the corresponding set $\mathcal{Z}^{(i,j)}$. From $e_1,e_2$ and $e_3$, the condition $X_{3,j} \neq 0$ forces at least one  $X_{i,j} \neq 0$ for $i=1,2$. Thus vectors in $\mathcal{S}$ with $X_{3,j} \neq 0$ also lie in $A$. Hence all vectors in $\mathcal{S}$ lie in $A$. Thus even in this case  $\alpha(\mathcal{M}_{\text{CFL}}, \mathbf{d}) \geq 6.$ From (\ref{eq: kappanek}), we get $\kappa(\mathcal{M}_{\text{CFL}}, \mathbf{d}) \leq 6$. Hence $\alpha(\mathcal{M}_{\text{CFL}})=\kappa(\mathcal{M}_{\text{CFL}})=6.$

Finally assume $N_e(\mathbf{d})=1$. Let $\mathbf{d}=(1,1,1).$ In addition to $e_1, e_2$ and $e_3$ consider the following set of equations
$ e_4: X_{2,1}=0, $
$ e_5: X_{2,2}=0, \text{ and }$
$ e_6: X_{2,3}=0.$
Let $\mathcal{S}$ be the subspace of $\mathbb{F}_q^9$, which consists of the vectors satisfying the set of equations $e_1,e_2, \ldots, e_6.$ We follow the similar argument as above to show that all  the vectors in $\mathcal{S}$ lie in $A$. By definition, $\mathbf{0}$ lies in $A$. All vector in $\mathcal{S}$ with $X_{1,j} \neq 0$ for $j=1,2,3$ are present in $A$. By $ e_4,e_5$ and $e_6$, all the vectors in $\mathcal{S}$ have $X_{2,j}=0$.
The condition $X_{3,j} \neq 0$ and the set of equations $e_1,e_2,\ldots,e_6$ force $X_{1,j} \neq 0$. Hence all vectors in $\mathcal{S}$ with $X_{3,j}\neq 0$ are present in $\mathcal{Z}^{(1,j)}$. Thus all vectors in $\mathcal{S}$ are present in $A$. Moreover, dim$(\mathcal{S}) \geq 9-6=3.$ Therefore $\alpha(\mathcal{M}_{\text{CFL}}, \mathbf{d}) \geq 3.$ From \eqref{eq: kappanek}, $\kappa(\mathcal{M}_{\text{CFL}}, \mathbf{d}) \leq 3$. Thus by \eqref{eq:alleqka}, $\alpha(\mathcal{M}_{\text{CFL}},\mathbf{d})=\kappa(\mathcal{M}_{\text{CFL}},\mathbf{d})=3.$
\label{eg:nekalp}
\end{exmp}

The example illustrates that the generalized independence number of the index coding problem $\mathcal{I}(\mathcal{M}_{\text{CFL}},\mathbf{d})$ is equal to its min-rank. For different demands,  the generalized index coding problem changes and for all those problems, min-rank and generalized independence number are shown to be equal. This can be shown for all values of $N$ as given in the theorem below.

\begin{thm}
	For $N=K$ and $M=1/N$, 
	\normalsize
	\small
\[
	\alpha(\mathcal{M}_{\text{CFL}}, \mathbf{d}) =  \kappa(\mathcal{M}_{\text{CFL}},\mathbf{d}) =
	\begin{cases}
	N (N_e(\mathbf{d}))              & \text{if} \  N_e(\mathbf{d}) \leq N-1\\
	N(N-1)             & \text{if} ~ N_e(\mathbf{d}) = N
	\end{cases},
\]	
\normalsize	
where $N_e(\mathbf{d})$ is the number of distinct demands. 
\label{Thm:NeK}
\end{thm}

\begin{IEEEproof}
In CFL prefetching scheme $\mathcal{M}_{\text{CFL}}$, each file $X_i, i \in [N]$ is split into $N$ subfiles $X_{i,1},X_{i,2},\ldots, X_{i,N}$. User $i, i \in [N]$ caches $Y_i=(X_{1,i} \oplus X_{2,i} \oplus \ldots \oplus X_{N,i})$. Let $X=(X_{1,1}, \ldots,  X_{1,N},X_{2,1}, \ldots, X_{2,N }, \ldots , X_{N,1}, \ldots, X_{N,N})$ be the vector obtained by concatenation of vectors $X_i, i \in [N]$ .

For a given demand $\mathbf{d}$, the delivery phase of the coded caching problem becomes a generalized index coding  problem $\mathcal{I}(\mathcal{M}_{\text{CFL}}, \mathbf{d})$ with $N^2$ messages and $N^2$ receivers.

First consider that all the demands are distinct, i.e., $N_e(\mathbf{d})=N$. Let the demand of the $i$th user be $X_{d_i}$. Thus $\mathbf{d}=(d_1,d_2, \ldots, d_N)$.
	Consider the set of $N$ equations denoted by $e_1, e_2, \ldots, e_N$, where 
	$$e_i: (X_{1,i} \oplus X_{2,i} \oplus \ldots \oplus X_{N,i}) =0.$$
	Let $\mathcal{S}$ be the subspace of $\mathbb{F}_q^{N^2}$ which consists of the vectors satisfying the set of equations $e_1,e_2,\ldots,e_N$. From rank nullity theorem,  we have dim$(\mathcal{S}) \geq N^2-N$.
	
	 For $\mathcal{I}(\mathcal{M}_{\text{CFL}}, \mathbf{d})$, from	 \eqref{eq:setz} we have,   $ \mathcal{Z}^{(i,j)} \triangleq \{Z \in \mathbb{F}_q^{N^2} : e_i, X_{d_i,j} \neq 0 \}.$ Let $A= \cup_{i,j \in [N]}\mathcal{Z}^{(i,j)} \cup \{\mathbf{0}\}$. The generalized independence number is the maximum dimension of any subspace of $\mathbb{F}_q^{N^2}$ in $A$. We show that $\mathcal{S}$ is such a subspace. For this we need to show that all vectors of $\mathcal{S}$ lie in $A$. By definition of $A$, the all zero vector $\mathbf{0}$ lies in $A$. Any other vector in $\mathcal{S}$ will have at least one non-zero coordinate. The vectors belonging to $\mathcal{S}$ having $X_{d_i,j} \neq 0$ belongs to the set  $\mathcal{Z}^{(i,j)}$. Thus all vectors in $\mathcal{S}$ lie in $A$. The generalized independence number $\alpha(\mathcal{M}_{\text{CFL}}, \mathbf{d}) \geq N^2-N.$ From \eqref{eq: kappanek}, we get $\kappa(\mathcal{M}_{\text{CFL}}, \mathbf{d}) \leq N^2-N$. Hence by \eqref{eq:alleqka}, we have $\alpha(\mathcal{M}_{\text{CFL}}, \mathbf{d})= \kappa(\mathcal{M}_{\text{CFL}}, \mathbf{d})=N^2-N.$
	
Consider the case where $N_e(\mathbf{d}) \leq N-1$. Without loss of generality we can assume that the first $N_e(\mathbf{d})$ users have distinct demands and that the $i$th user demands the file $X_{d_i}$ for $i \in  [N_e(\mathbf{d})]$. Without loss of generality, we can assume that the set of indices of the files that are not demanded are $ N_e(\mathbf{d})+1, N_e(\mathbf{d})+2, \ldots, N$. There are $U=N-N_e(\mathbf{d})$ files which are not demanded. In addition to $e_1,e_2, \ldots, e_N$, consider the following set of equations  $ X_{N_e(\mathbf{d})+i,j}=0$, for $i \in [U-1], j \in [N]$. The number of equations is thus $N+ N(U-1)=NU=N(N-N_e(\mathbf{d}))$. Let $\mathcal{S}$ be the subspace of $\mathbb{F}_q^{N^2}$ which consists of vectors satisfying these equations. Hence, dim$(\mathcal{S}) \geq N^2-N(N-N_e(\mathbf{d})) = N(N_e(\mathbf{d})).$ By definition, $\mathbf{0}$ lies in $A$. Any vector with the coordinate $X_{d_i,j} \neq 0$ for $i \in [N_e(\mathbf{d})]$ lies in $\mathcal{Z}^{(i,j)}$. The set of equations force all $X_{i,j}=0$ for $i \in \{N_e(\mathbf{d}), \ldots, N-1\}$. Moreover if $X_{N,j} \neq 0$ the set of equations force some $X_{d_i,j} \neq 0$ for some $i \in [N_e(\mathbf{d})]$. Hence any vector with $X_{N,j} \neq 0$ lies in  some $\mathcal{Z}^{(i,j)}$ for $i \in [N_e(\mathbf{d})]$. Thus all vectors in $\mathcal{S}$ lie in $A$.
	Therefore $\alpha(\mathcal{M}_{\text{CFL}}, \mathbf{d}) \geq \text{dim}(\mathcal{S}) \geq N(N_e(\mathbf{d})).$ Applying (\ref{eq: kappanek}) and (\ref{eq:alleqka}), $\alpha(\mathcal{M}_{\text{CFL}},\mathbf{d})=\kappa(\mathcal{M}_{\text{CFL}},\mathbf{d})=N(N_e(\mathbf{d})).$
	
\end{IEEEproof}

\subsection{Number of users more than the number of files ($K >N$)}
\label{Sec:KgreaterthanN}

In the CFL prefetching scheme for $N < K$, each file is split into $NK$ subfiles. Hence the number of messages in  $\mathcal{I}(\mathcal{M}_{\text{CFL}}, \mathbf{d})$ is $N^2K$. Each user is split into $NK$ receivers in  $\mathcal{I}(\mathcal{M}_{\text{CFL}}, \mathbf{d})$ each demanding a single message. Thus there are a total of $NK^2$ receivers. From the expressions for achievable rates in \cite{CFL}, we get the min-rank $\kappa(\mathcal{M}_{\text{CFL}}, \mathbf{d})$ as
\begin{equation}
\kappa(\mathcal{M}_{\text{CFL}}, \mathbf{d}) \leq 
\begin{cases}
NK (N_e(\mathbf{d}))              & \text{if} \  N_e(\mathbf{d}) \leq N-1\\
N^2(K-1)             & \text{if } N_e(\mathbf{d})= N
\end{cases}.
\label{eq: kappannk}
\end{equation}
 We find the generalized independence number $\alpha(\mathcal{M}_{\text{CFL}}, \mathbf{d})$ for $\mathcal{I}(\mathcal{M}_{\text{CFL}}, \mathbf{d})$. The technique of obtaining $\alpha(\mathcal{M}_{\text{CFL}}, \mathbf{d})$  is illustrated in the following example.
 
 \begin{exmp}
 	Consider a coded caching problem with    $N=3$, $K=4$ and $M=1/4$. According to the CFL scheme each file is split into $NK=12$ subfiles as 
 	$ X_1=(X_{1,1}, X_{1,2}, \ldots,  X_{1,12})$,  
 	$ X_2=(X_{2,1}, X_{2,2}, \ldots, X_{2,12})$ and 	$ X_3=(X_{3,1}, X_{3,2}, \ldots, X_{3,12})$. Let $X=(X_{1,1}, \ldots, X_{1,12}, \ldots, X_{3,1}, \ldots, X_{3,12})$ denote the vector obtained by concatenating $X_1, X_2$ and $X_3$.
 The cache of the $i$th user contains three coded packets $Y_i=(X_{1,3(i-1)+j} \oplus X_{2,3(i-1)+j} \oplus X_{3,3(i-1)+j})$ for $j=1,2,3.$ 
For a given demand $\mathbf{d}$, this problem becomes a generalized index coding problem $\mathcal{I}(\mathcal{M}_{\text{CFL}}, \mathbf{d})$ having 36 messages and 48 receivers. 

First consider that i.e., $N_e(\mathbf{d})=N=3$ and $d=(1,2,3,1).$
 Consider the equations given by $e_{i,j}: (X_{1,3(i-1)+j} \oplus X_{2,3(i-1+j)} \oplus X_{3,3(i-1)+j})=0$ for $i=1,2,3$ and $j=1,2,3$. Thus there are nine equations. Let $\mathcal{S}$ be the subspace of vectors in $\mathbb{F}_q^{36}$ satisfying these nine equations. From rank nullity theorem, we get dim$(\mathcal{S}) \geq 36-9=27$. For this case, (\ref{eq:setz}) can be rewritten as
 	$ \mathcal{Z}^{(i,j)} \triangleq \{Z \in \mathbb{F}_q^{36} : e_{i,1},e_{i,2},e_{i,3}, X_{d_i,j} \neq 0 \}$ for $i \in [4]$. Let $A= \cup_{i \in [4],j \in [12]}\mathcal{Z}^{(i,j)} \cup \{\mathbf{0}\}$. The generalized independence number is the maximum dimension of any subspace of $\mathbb{F}_q^{36}$ in $A$. We claim that $\mathcal{S}$ is such a subspace. This would mean that $\alpha(\mathcal{M}_{\text{CFL}}, \mathbf{d}) \geq \text{dim}(\mathcal{S}) \geq 27$. For this we need to show that all vectors in $\mathcal{S}$ lie in $A$. By definition of $A$, the all zero vector $\mathbf{0}$ lies in $A$. Any other vector in $\mathcal{S}$ will have at least one non-zero coordinate. All vectors in $\mathcal{S}$, having $X_{d_i,j} \neq 0$ belongs to  $\mathcal{Z}^{(i,j)}$. Thus all vectors in $\mathcal{S}$ lie in $A$ and $\alpha(\mathcal{M}_{\text{CFL}}, \mathbf{d}) \geq 27.$ From (\ref{eq: kappannk}), we get $\kappa(\mathcal{M}_{\text{CFL}}, \mathbf{d}) \leq 3^2(4-1)=27$. Hence by (\ref{eq:alleqka}), we have $\alpha(\mathcal{M}_{\text{CFL}})=\kappa(\mathcal{M}_{\text{CFL}})=27.$
 	
 	Consider now that $N_e(\mathbf{d})=2$ and $\mathbf{d}=(1,2,1,2)$. In addition to the nine equations $e_{i,j}$ for $i=1,2,3$ and $j=1,2,3$, consider three more equations $e_{4,j}: (X_{1,9+j} \oplus X_{2,9+j)} \oplus X_{3,9+j})=0$ for $j=1,2,3$. Thus we consider a set of twelve equations given by $E =\{e_{i,j}: i \in [4], j \in [3]\}$. Let $\mathcal{S}$ be the subspace of $\mathbb{F}_q^{36}$ consisting of vectors which satisfy the equations in $E$. Hence from rank nullity theorem, we have dim$(\mathcal{S}) \geq 36-12 =24.$ By definition, $\mathbf{0}$ lies in $A$. Any non-zero vector in $\mathcal{S}$ with $X_{d_i,j} \neq 0$ for $i=1,2$ lies in the corresponding $\mathcal{Z}^{(i,j)}$. By $E$, any $X_{3,j} \neq 0$ forces some $X_{i,j} \neq 0 $ for $i=1,2$ and hence such vectors also lie in $A$. Thus all vectors in $\mathcal{S}$ lie in $A$. Therefore  $\alpha(\mathcal{M}_{\text{CFL}}, \mathbf{d}) \geq \text{dim} (\mathcal{S}) \geq  24.$  From (\ref{eq: kappannk}), we get $\kappa(\mathcal{M}_{\text{CFL}}, \mathbf{d}) \leq 12(2)=24$. Hence by (\ref{eq:alleqka}), we have $\alpha(\mathcal{M}_{\text{CFL}})=\kappa(\mathcal{M}_{\text{CFL}})=24.$
 	
 	Finally consider $N_e(\mathbf{d})=1$ and $\mathbf{d} =(1,1,1,1)$. The files $X_2$ and $X_3$ are not demanded by any user. In addition to the equations in $E$, here we consider a set of equations $X_{2,j}=0$ for $j \in [12]$. Thus there are 24 equations in total. Let $\mathcal{S}$ be the subspace of $\mathbb{F}_q^{36}$ which satisfy these equations. By rank nullity theorem, the dimension of $\mathcal{S}$ is given by dim$(\mathcal{S}) \geq 36-24=12.$ The next step is to show that all the vectors in $\mathcal{S}$ lie in $A$. The all zero vector $\mathbf{0}$ lies in $A$ by definition. Any non-zero vector in $\mathcal{S}$ with $X_{1,j} \neq 0$ for $j \in [12]$ lies in the corresponding $\mathcal{Z}^{(i,j)}$. From the set of equations, we have $X_{2,j}=0$ for $j \in [12]$.	
 	By $E$, any $X_{3,j} \neq 0$ forces $X_{1,j} \neq 0 $ for $j \in [12]$ and hence such vectors also lie in $A$. Thus all vectors in $\mathcal{S}$ lie in $A$. Therefore  $\alpha(\mathcal{M}_{\text{CFL}}, \mathbf{d}) \geq \text{dim} (\mathcal{S}) \geq  12.$  From \eqref{eq: kappannk}, we get $\kappa(\mathcal{M}_{\text{CFL}}, \mathbf{d}) \leq 12(1)=12$. Hence by \eqref{eq:alleqka}, we have $\alpha(\mathcal{M}_{\text{CFL}})=\kappa(\mathcal{M}_{\text{CFL}})=12.$ 
 	\label{eg:nnkalp}
 \end{exmp}
The theorem below gives the expression for 
$\alpha(\mathcal{M}_{\text{CFL}},\mathbf{d})$, when $N < K$.

\begin{thm}
	For $N < K$ and $M=1/K$, 
	\normalsize
	\small
\[
	\alpha(\mathcal{M}_{\text{CFL}}, \mathbf{d}) = \kappa(\mathcal{M}_{\text{CFL}}, \mathbf{d}) =
	\begin{cases}
	NK (N_e(\mathbf{d}))              & \text{if} \  N_e(\mathbf{d}) \leq N-1\\
	N^2(K-1)             & \text{if } N_e(\mathbf{d})=N,
	\end{cases}
\]
\normalsize
	where $N_e(\mathbf{d})$ is the number of distinct demands.  \label{Thm:NnK}	
\end{thm}
\begin{IEEEproof}
	For $N < K$ and $M=1/K$, the CFL prefetching scheme $\mathcal{M}_{\text{CFL}}$ is as follows. Each file is split into $NK$ subfiles $X_i=(X_{i,1}, X_{i,2}, \ldots, X_{i,NK})$. User $i, i \in [K]$ caches $N$ coded packets given by $Y_i=X_{1,N(i-1)+j} \oplus \ldots \oplus X_{N,N(i-1)+j},$ for $j \in [N].$ Let $X=(X_{1,1}, \ldots, X_{1,NK}, \ldots, X_{N,1}, \ldots, X_{N,NK})$ be the vector obtained by the concatenation of vectors $X_i, i \in [N].$
For a given demand $\mathbf{d}$, this problem becomes a generalized index coding problem $\mathcal{I}(\mathcal{M}_{\text{CFL}}, \mathbf{d})$ with $N^2K$ messages and $NK^2$ receivers. 

First consider that all the demands are distinct, i.e., $N_e(\mathbf{d})=N$. Without loss of generality we can assume  that the first $N$ users demand distinct files such that the $i$th user demands $X_{d_i}$ for $i=1,2,\ldots, N$. Thus $\mathbf{d}=(d_1,d_2, \ldots, d_K)$ such that $d_i \neq d_j$ for $i,j \in [N]$. 
Let $E =\{e_{i,j}: i \in [K], j \in [N]\}$ represent a set of $NK$ equations, where $e_{i,j}: (X_{1,N(i-1)+j} \oplus X_{2,N(i-1)+j} \oplus \ldots \oplus X_{N,N(i-1)+j})=0$.
We consider a subset of the equations in $E$ of the form $e_{i,j}$ for $i,j \in [N]$. There are $N^2$ such equations. Let $\mathcal{S}$ be the subspace of $\mathbb{F}_q^{N^2K}$ consisting of vectors satisfying these equations. From rank nullity theorem we have dim$(\mathcal{S}) \geq N^2K-N^2=N^2(K-1)$.
	
	For  $\mathcal{I}(\mathcal{M}_{\text{CFL}}, \mathbf{d})$, \eqref{eq:setz} can be rewritten as $ \mathcal{Z}^{(i,j)} \triangleq \{Z \in \mathbb{F}_q^{N^2K} : e_{i,k} \text{ for } k \in [N], X_{d_i,j} \neq 0 \}$ for $i \in [K]$ and $j \in [NK]$. Let $A= \cup_{i \in [K],j \in [NK]}\mathcal{Z}^{(i,j)} \cup \{\mathbf{0}\}$. The generalized independence number is the maximum dimension of any subspace of $\mathbb{F}_q^{N^2K}$ in $A$. We show that $\mathcal{S}$ is such a subspace. For this we need to show that all vectors of $\mathcal{S}$ lie in $A$. By definition of $A$, the all zero vector $\mathbf{0}$ lies in $A$. 
The vectors belonging to $\mathcal{S}$ having 	$X_{d_i,j} \neq 0$ belongs to the set $\mathcal{Z}^{(i,j)}$.
Thus all vectors in $\mathcal{S}$ lie in $A$ and $\alpha(\mathcal{M}_{\text{CFL}}, \mathbf{d}) \geq N^2(K-1).$ From \eqref{eq: kappannk}, we get $ \kappa(\mathcal{M}_{\text{CFL}}) \leq N^2(K-1)$. Hence by \eqref{eq:alleqka}, we have $\alpha(\mathcal{M}_{\text{CFL}})=\kappa(\mathcal{M}_{\text{CFL}})=N^2(K-1).$

Consider the case where $N_e(\mathbf{d}) \leq N-1$. Let the first $N_e(\mathbf{d})$ demands be distinct and the $i$th user demands $X_{d_i}$ for $i \in [N_e(\mathbf{d})].$ Without loss of generality we can assume that the indices of the files which are not demanded are $N_e(\mathbf{d})+1, \ldots, N$. There are $U=N-N_e(\mathbf{d})$ files which are not demanded.
In addition to the $NK$ equations in $E$, 
	 consider the following equations $ X_{N_e(\mathbf{d})+i,j}=0$, for $i \in  [U-1]$ and $j \in [NK]$. The number of equations is thus $NK+ NK(U-1)=NKU=NK(N-N_e(\mathbf{d}))$. Let $\mathcal{S}$ be the subspace of $\mathbb{F}_q^{N^2K}$ which consists of the vectors satisfying these equations. By rank nullity theorem,  dim$(\mathcal{S}) \geq N^2K-NK(N-N_e(\mathbf{d})) = NK(N_e(\mathbf{d})).$ By definition, $\mathbf{0}$ lies in $A$. Any vector in $\mathcal{S}$ with the coordinate $X_{d_i,j} \neq 0$ for $i \in [N_e(\mathbf{d})]$ lies in $\mathcal{Z}^{(i,j)}$. The set of equations force all $X_{i,j}=0$ for $i \in \{N_e(\mathbf{d}), \ldots, N-1\}$ and $j \in [NK]$. Moreover by the set of equations in $E$, $X_{N,j} \neq 0$ would mean some other $X_{d_i,j} \neq 0$ for $i \in [N_e(\mathbf{d})]$. Hence any vector with $X_{N,j} \neq 0$ lies in  some $\mathcal{Z}^{(i,j)}$ for $i \in [N_e(\mathbf{d})]$. Thus all vectors in $\mathcal{S}$ lie in $A$.
	Therefore $\alpha(\mathcal{M}_{\text{CFL}}, \mathbf{d}) \geq \text{dim}(\mathcal{S}) \geq NK(N_e(\mathbf{d})).$ From \eqref{eq: kappanek}, we have $\kappa(\mathcal{M}_{\text{CFL}},\mathbf{d}) \leq NK(N_e(\mathbf{d}))$. Hence from \eqref{eq:alleqka}, $\alpha(\mathcal{M}_{\text{CFL}},\mathbf{d})=\kappa(\mathcal{M}_{\text{CFL}},\mathbf{d})=NK(N_e(\mathbf{d})).$	
	
\end{IEEEproof}

\section{Optimal Error Correcting Delivery Scheme for CFL Prefetching Scheme}
\label{Sec:ECDforCFL}

In this section we give an expression for the average rate and worst case rate for a $\delta$-error correcting delivery scheme for CFL prefetching scheme. Also we propose a $\delta$-error correcting delivery scheme for this case.
From Theorem \ref{Thm:NeK} and Theorem \ref{Thm:NnK}, we can conclude that for all the generalized index coding problems $\mathcal{I}(\mathcal{M}_{\text{CFL}},\mathbf{d})$ induced from the CFL prefetching scheme,
\begin{equation}
\alpha(\mathcal{M}_{\text{CFL}},\mathbf{d}) = \kappa(\mathcal{M}_{\text{CFL}},\mathbf{d}).
\label{eq:alpekap}
\end{equation}
Hence, the $\alpha$ and $\kappa$ bounds in \eqref{eq:bds} meet. Using this the optimal error correcting delivery scheme can be constructed for  CFL prefetching scheme and hence the average rate can be calculated as given in the following theorem.

\begin{thm}
	For a coded caching problem with CFL prefetching scheme for $M=1/K$,
	$$ R^*(\mathcal{M}_{\text{CFL}}, \delta) = \mathbb{E}_{\mathbf{d}}\bigg[\frac{N_q[\kappa({\mathcal{M}_{\text{CFL}}, \mathbf{d}}), 2\delta+1]}{{n_{\text{CFL}}}}\bigg],$$
	where $n_{\text{CFL}}$ is the number of subfiles into which each file is divided in the CFL scheme.
 	Furthermore, for $ M \in [0,1/K]$, $R^*(\mathcal{M}_{\text{CFL}}, \delta)$ equals the lower convex envelope of its values at $M=0$ and $M=1/K$.
\end{thm}
\begin{IEEEproof}
From (\ref{eq:alpekap}) and (\ref{eq:bds}), we can conclude that for any generalized index coding problem induced from the coded caching problem with CFL prefetching, the $\alpha$ and $\kappa$ bounds meet. Thus the optimal error correcting delivery scheme would be the concatenation of the CFL delivery scheme with an optimal linear error correcting code. The optimal length or equivalently the optimal number of transmissions required for $\delta$ error corrections in those generalized index coding problems is thus $N_q[\kappa({\mathcal{M}_{\text{CFL}}, \mathbf{d}}), 2\delta+1]$ and hence the statement of the theorem follows for $M=1/K$. 
For  $M \in [0,1/K]$, the lower convex envelope of values of $R^*(\mathcal{M}_{\text{CFL}}, \delta)$ is achieved by using memory sharing.
\end{IEEEproof}

%%%%%%%%%%%%%%%%%%%%%%%%
\begin{cor}
	For a coded caching problem with CFL prefetching scheme for $M=1/K$,
$$ R^*_{\text{worst}}(\mathcal{M}_{\text{CFL}}, \delta) = \frac{N_q[\kappa_{\{\text{worst}\}}({\mathcal{M}_{\text{CFL}}, \mathbf{d}}), 2\delta+1]}{{n_{\text{CFL}}}},$$
where the value of  $\kappa_{\{\text{worst}\}}({\mathcal{M}_{\text{CFL}}, \mathbf{d}})$ is obtained from (\ref{eq: kappanek}) and (\ref{eq: kappannk}) when $N_e(\mathbf{d})=N$. 
 	Furthermore, for $ M \in [0,1/K]$, $R^*_{\text{worst}}(\mathcal{M}_{\text{CFL}}, \delta)$  equals the lower convex envelope of its values at $M=0$ and $M=1/K$.
	
\end{cor}
\begin{IEEEproof}
	Worst case rate is required when the number of distinct demands is maximum. This happens when $N_e({\mathbf{d}})= N.$
\end{IEEEproof}

Since the $\alpha$ and $\kappa$ bounds become equal for $\mathcal{I}(\mathcal{M}_{\text{CFL}},\mathbf{d})$, the optimal coded caching delivery scheme here would be the concatenation of the CFL delivery scheme with optimal classical error correcting scheme which corrects $\delta$ errors. Decoding can be done by syndrome decoding for error correcting generalized index codes proposed in \cite{DSC,ByC}.

In the remaining part of this section, few examples of optimal error correcting delivery scheme for coded caching problems with CFL prefetching are given.

\begin{exmp}
Consider the coded caching problem considered in Example \ref{eg:nekalp}. First consider that $N_e(\mathbf{d})=3$ and $\mathbf{d}=(1,2,3)$. We have shown that for this case $\kappa({\mathcal{M}_{\text{CFL}}, \mathbf{d}})=6$. The transmissions in the CFL scheme are $T_1: X_{2,1}$, $T_2: X_{3,1}$, $T_3: X_{1,2}$, $T_4: X_{3,2}$, $T_5: X_{1,3}$ and $T_6: X_{2,3}$. If $\delta=1$ transmission error needs to be corrected, then from \cite{Gra}, we have $N_2[6,3]=10$. A generator matrix corresponding to $[10,6,3]_2$ code is 
$$
	\bf{G}=
	\begin{bmatrix}
	1 & 0 & 0 & 0 & 0 & 0 & 1 & 1 & 0 & 0 \\
	0 & 1 & 0 & 0 & 0 & 0 & 1 & 0 & 1 & 0 \\
	0 & 0 & 1 & 0 & 0 & 0 & 1 & 0 & 0 & 1 \\
	0 & 0 & 0 & 1 & 0 & 0 & 0 & 1 & 1 & 0 \\
	0 & 0 & 0 & 0 & 1 & 0 & 0 & 1 & 0 & 1 \\
	0 & 0 & 0 & 0 & 0 & 1 & 0 & 0 & 1 & 1 
	\end{bmatrix}.
	$$
The optimal single error correcting delivery scheme is the concatenation of the CFL delivery scheme with the above code. Thus single error correcting delivery scheme involves 10 transmissions. In addition to $T_1, \ldots, T_6$ the following transmissions are required.
$$ T_7: X_{2,1} \oplus X_{3,1} \oplus X_{1,2},$$
$$ T_8: X_{2,1} \oplus X_{3,2} \oplus X_{1,3},$$   
$$ T_9: X_{3,1} \oplus X_{3,2} \oplus X_{2,3} \text{ and }$$
$$ T_{10}: X_{1,2} \oplus X_{1,3} \oplus X_{2,3}.$$

Now consider $N_e(\mathbf{d})=2$ and $\mathbf{d}=(1,2,1)$. Even for this case $\kappa({\mathcal{M}_{\text{CFL}}, \mathbf{d}})=6$. The transmissions in the CFL scheme are $T_1: X_{1,1}$, $T_2: X_{1,2}$, $T_3: X_{1,3}$, $T_4: X_{2,1}$, $T_5: X_{2,2}$ and $T_6: X_{2,3}$. For single error correction,  the concatenation is done with the same $[10,6,3]_2$ code. Considering the same generator matrix as before, the additional transmissions in the error correcting delivery scheme are
   $$ T_7: X_{1,1} \oplus X_{1,2} \oplus X_{1,3},$$
   $$ T_8: X_{1,1} \oplus X_{2,1} \oplus X_{2,3},$$   
   $$ T_9: X_{1,2} \oplus X_{2,1} \oplus X_{2,3} \text{ and }$$
   $$ T_{10}: X_{1,3} \oplus X_{2,2} \oplus X_{2,3}.$$
   
Finally consider $N_e(\mathbf{d})=1$ and $\mathbf{d}=(1,1,1)$. For this case, $\kappa({\mathcal{M}_{\text{CFL}}, \mathbf{d}})=3$. The CFL transmission scheme involves the following three transmissions $T_1: X_{1,1}$, $T_2: X_{1,2}$ and $T_3: X_{1,3}$. For single error correction, we have from \cite{Gra} that $N_2[3,3]=6$. A generator matrix for the $[6,3,3]_2$ code is
$$
	\bf{G}=
	\begin{bmatrix}
	1 & 0 & 0 & 1 & 1 & 0 \\
	0 & 1 & 0 & 1 & 0 & 1 \\
	0 & 0 & 1 & 0 & 1 & 1
	\end{bmatrix}.
	$$
The optimal single error correcting delivery scheme is thus the concatenation of the CFL delivery scheme with the above code. The additional transmissions required apart from $T_1,T_2$ and $T_3$ are
 $$ T_4: X_{1,1} \oplus X_{1,2},$$
 $$ T_5: X_{1,1} \oplus X_{1,3} \text{ and }$$
  $$ T_6: X_{1,2} \oplus X_{1,3}.$$
  
Decoding is done by syndrome decoding for generalized index codes \cite{DSC} \cite{ByC}.

\end{exmp}

\begin{exmp}
	Consider the coded caching problem considered in Example \ref{eg:nnkalp}. Consider that $N_e(\mathbf{d})=3$ and $\mathbf{d}=(1,2,3,1)$. We have shown that for this case $\kappa({\mathcal{M}_{\text{CFL}}, \mathbf{d}})=27$. The transmissions in the CFL scheme are $T_1: X_{2,1}$, $T_2: X_{3,1}$, $T_3: X_{2,2}$, $T_4: X_{3,2}$, $T_5: X_{2,3}$, $T_6: X_{3,3}$, $T_7: X_{1,4}$, $T_8: X_{3,4}$, $T_9: X_{1,5}$, $T_{10}: X_{3,5}$, $T_{11}: X_{1,6}$, $T_{12}: X_{3,6}$, $T_{13}: X_{1,7}$, $T_{14}: X_{2,7}$, $T_{15}: X_{1,8}$, $T_{16}: X_{2,8}$, $T_{17}: X_{1,9}$, $T_{18}: X_{2,9}$, $T_{19}: X_{2,10}$, $T_{20}: X_{3,10}$, $T_{21}: X_{2,11}$, $T_{22}: X_{3,11}$, $T_{23}: X_{2,12}$, $T_{24}: X_{3,12}$, $T_{25}: X_{1,1} \oplus X_{1,10}$, $T_{26}: X_{1,2} \oplus X_{1,11}$ and $T_{27}: X_{1,3} \oplus X_{1,12}$. If $\delta=1$ transmission error needs to be corrected, then from \cite{Gra}, we have $N_2[27,3]=42$. Consider $\delta=1$ transmission error to be corrected. The optimal single error correcting delivery scheme involves concatenation of CFL delivery scheme with a generator matrix corresponding to the $[42,27,3]_2$ code.
	
\end{exmp}

\section{Conclusion}
In this work, we obtained the minimum number of transmissions required for a $\delta$-error correcting delivery scheme for coded caching problems with the CFL prefetching scheme. We proposed an optimal error correcting delivery scheme for the above case. We also found closed form expressions for the average rate and the peak rate for these problems.

%%%%%%%%%%%%%%%%%%%%%%%%%%%%%%%%%%%%%%%%%%%%%%%%%%%%%%%%%%%%%%%%%%%%%
\section*{Acknowledgment}
This work was supported partly by the Science and Engineering Research Board (SERB) of Department of Science and Technology (DST), Government of India, through J.C. Bose National Fellowship to B. Sundar Rajan.

\end{document}